\documentclass[preprint,showpacs,preprintnumbers,amsmath,amssymb,prl]{revtex4}

\usepackage{graphicx}
\usepackage{dcolumn}
\usepackage{bm}
\usepackage{epsfig}

\begin{document}

\preprint{TRINLAT-03/02,Edinburgh 2003/10}

\title{The spectrum of $D_s$ mesons from lattice QCD}%
\author{UKQCD Collaboration}
\noaffiliation
\author{A.~Dougall}
\affiliation{School of Mathematics, Trinity College, Dublin 2, Ireland}
\author{R.D.~Kenway}
\author{C.M.~Maynard\footnote{Author for correspondence, cmaynard@ph.ed.ac.uk} }
\affiliation{School of Physics, University of Edinburgh, Edinburgh, EH9 3JZ, 
UK}
\author{C.~McNeile}
\affiliation{
Theoretical Physics Division, Dept. of Mathematical Sciences,
          University of Liverpool, Liverpool L69 3BX, UK}

\date{\today}

\begin{abstract}
The spectrum of orbitally excited $D_s$ mesons is computed in the
continuum limit of quenched lattice QCD. The results are consistent
with the interpretation that the narrow resonance in the $D_s \pi^0$
channel discovered by the BABAR Collaboration is a $J^P=0^+$
$c\bar{s}$ meson. Furthermore, within statistical errors, the
$1^+-1^-$ and the $0^+-0^-$ mass splittings are equal, in agreement
with the chiral multiplet structure predicted by heavy hadron chiral
effective theory. On our coarsest lattice we present results from the
first study of orbitally excited $D_s$ mesons with two flavors of
dynamical quarks, with mass slightly larger than the strange quark
mass. These results are consistent with the quenched data.

\end{abstract}

\pacs{12.38.Gc,14.40.Lb}

\maketitle

\section{\label{sec:intro}Introduction}
The recent discovery by the BABAR collaboration of a new resonance,
with a mass around $2.32$ GeV and a narrow width, in the $D_s^+ \pi^0$
final state~\cite{BaBar_Ds} has provoked a great deal of interest from
experimenters and theorists alike. The CLEO
collaboration~\cite{CLEO_Ds} has confirmed this resonance. Both
experiments interpret this as the lowest lying of the four $P$-wave
states, the ${}^3P_0$ with $J^P=0^+$. A variety of theory papers have
been published~\cite{cahn_Ds,barnes_Ds,van_bereven_Ds,cheng_Ds,bardeen_Ds,szczepaniak_Ds,colangelo_Ds,bali_Ds,Terasaki_Ds,Dai_Ds}
either supporting this interpretation or presenting alternative
hypotheses of ``exotic'' states. These alternatives are motivated by
potential  model results~\cite{gi_quark_model,gk_quark_model}, which
suggest that the $c\bar{s}$ scalar meson mass is around $2.48$ GeV and hence
above the $DK$ threshold.  One could argue the merits of
a particular model, but this becomes irrelevant if the spectrum can
directly determined from QCD. A mass below the $DK$ threshold would also
explain the narrowness of the state.

There have been a number of previous lattice QCD calculations of the
$L=1$ states in the $D_s$ meson spectrum.  Hein {\em et
al.}~\cite{Hein:2000qu} obtained 500(80) MeV for the $D^{\star}_{s0}
-D_s$ mass splitting (from their figure 27) using NRQCD at a fixed
lattice spacing of 0.18 fm.  Lewis and Woloshyn~\cite{Lewis:2000sv}
obtained 530(15)(5) MeV at a fixed lattice spacing of 0.11 fm also
using NRQCD.  With quenched relativistic heavy quarks,
Boyle~\cite{Boyle:1997aq} obtained a mass splitting of 544(20) MeV at
a fixed lattice spacing of 0.07 fm. Although the calculation used the
same ensemble of gauge configurations as this work, the propagators
were computed using a slightly different action and different
definition of the lattice quark mass.  Recently Bali~\cite{bali_Ds}
presented results in the static limit for the heavy quark and obtained
a value of $468(43)(24)$ MeV for the scalar-pseudoscalar mass
splitting in this limit.

All the previous lattice QCD calculations of the $D_s$ spectrum were
done at fixed lattice spacing in quenched QCD. In this calculation we
take the continuum limit in quenched QCD, so that lattice artifacts
are under control.  The lattice volumes are large enough (greater than
$1.5^3 \ {\rm fm}^3$) that finite size effects should be small.  We also
report results from the first unquenched calculation at fixed lattice
spacing.  The remaining systematic uncertainty is due to dynamical $u$
and $d$ quarks having unphysically large masses.

In the heavy quark limit, the spin of the heavy quark decouples from
the rest of the system.  The observable states can be labelled by the
total angular momentum of the light quark $j$
\cite{bartelt_D_spec}. The $P$-wave states have $L=1$ and thus
$j=\{\frac{1}{2},\frac{3}{2}\}$.  This combined with the spin of the
heavy quark produces two doublets. The $j=\frac{3}{2}$ doublet
contains a $J=2$ and a $J=1$ state, the $j=\frac{1}{2}$ doublet
contains a $J=1$ and $J=0$ state. The two $J=1$ states do not have
definite charge conjugation and so can mix.  On the lattice only
the lightest state in this channel can be determined easily.

In the double limit of heavy quark and chiral symmetry, the two heavy
light multiplets, $\{0^-,1^-\}$ and $\{0^+,1^+\}$, are degenerate. The
effect of spontaneous chiral symmetry breaking is to split these
parity partners, such that the mass splittings $1^+-1^-$ and $0^+-0^-$
are equal~\cite{bardeen_Ds}. This is confirmed by the CLEO
Collaboration~\cite{CLEO_Ds} who obtain splittings of $351(2)$ and
$350(1)$ MeV respectively. It is interesting to explore the extent to
which QCD reproduces this remarkable agreement.

\section{Lattice details}
The spectrum of $c\bar{s}$ mesons has been determined on four
ensembles of gauge configurations. Three have different lattice
spacings ($a$) and were generated in the quenched approximation, which
enables the continuum limit to be taken.  The fourth ensemble was
generated with two degenerate flavours of dynamical quarks, and the
lattice spacing matched to that of the coarsest quenched ensemble. For
all the ensembles, the Wilson gauge action and the non-perturbatively
${\mathcal O}(a)$ improved Wilson fermion action were used. The
lattice parameters are detailed in Table~\ref{tab:ensembles}.  The
procedures for generating the dynamical ensemble and matching to the
coarsest quenched ensemble are described in~\cite{DFLHS}.

Meson correlation functions were computed with several different heavy
quark masses which span the charm quark mass, and several light quark
masses around the strange quark mass.  For the dynamical ensemble only
one sea quark mass is used, for which the ratio $m_{PS}/m_V=0.70(1)$
when $m_{\rm sea}=m_{\rm valence}$. This corresponds to QCD with two
dynamical flavours of mass slightly above the strange quark mass. The
details of extracting the spectrum from lattice correlation functions,
and the results for the $S$-wave $D$ meson spectrum for the finer two
lattice spacings can be found in~\cite{np_imp_fb}. To measure the mass
splittings, the ratio of correlation functions at large times is
fitted. We have checked that the results from computing the $1^+-0^+$
mass splitting directly is the same, with the same statistical errors
as obtained by combining the results obtained for the $1^+-1^-$,
$0^+-0^-$ and $1^--0^-$ splittings.

In quenched QCD, or in simulations with unphysical heavy sea quarks,
there is an ambiguity in the determination of the lattice spacing in
physical units. For typical simulation parameters used today this is
estimated to be of the order of $10\%$.  The scale is set throughout
this calculation from the static quark potential using
$r_0$~\cite{sommer_r0,wittig_r0}. The value of $r_0/a$ is unambiguous
for each ensemble, and so is a good choice for comparing results from
different ensembles. However, there is no agreed experimental value
for $r_0$. Sommer originally advocated $r_0=0.5$~fm. 
The lattice spacing obtained from the $K^{\star}/K$ mass ratio (method of
planes)~\cite{Allton:1997yv} corresponds to $r_0 \sim 0.55$~fm 
on the ensembles used in this work. Determinations of
the lattice spacing from the kaon decay
constant, the nucleon mass or the rho mass correspond to $r_0$ values
ranging from approximately $0.5$ to $0.55$ fm 
\cite{Garden:1999fg,Aoki:2002uc,DFLHS,Becirevic:1998ua,np_imp_fb}.
For these reasons we take the value of $r_0$ to be 0.55~fm. The analysis
has been repeated using $r_0=0.5$~fm throughout, and the difference
is taken as an estimate of the systematic uncertainty in the scale.

The strange quark mass is set by from the light-light pseudoscalar
mass with the experimental kaon mass as input~\cite{Hepburn:2002wa}.
Similarly the charm quark mass is set from the heavy-light
pseudoscalar mass with the experimental $D_s$ meson mass as input.

\section{Results}
\begin{figure}
\begin{center}
\epsfig{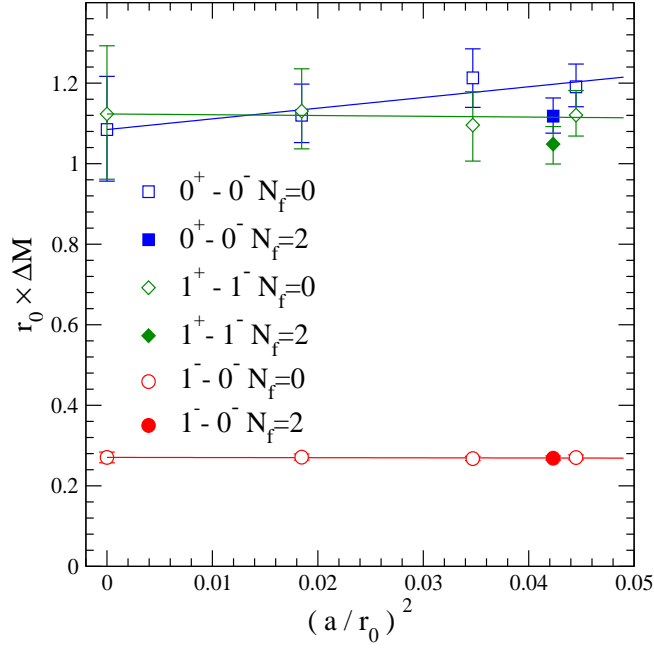} 
\caption{\label{fig:cont_extrap}Continuum extrapolation of quenched 
results (open symbols) and $N_f=2$ results in dimensionless units at
fixed lattice spacing (solid symbols, offset for clarity ) for mass
splittings in the $D_s$ system.}
\end{center}
\end{figure}

For each of the four ensembles, the mass splittings $1^+-1^-$ and
$0^+-0^-$ are equal within statistical errors. These results and the
continuum extrapolation which is linear in $a^2$ are shown in
figure~\ref{fig:cont_extrap}.  Furthermore, these splittings are also
equal in the continuum limit, in agreement with heavy hadron chiral
effective theory and experiment. At the coarsest lattice spacing the
effect of introducing sea quarks with a mass close to the strange is
to slightly lower the $1^+-1^-$ and the $0^+-0^-$ mass splittings, but
this is not statistically significant.

\begin{figure}
\begin{center}
\epsfig{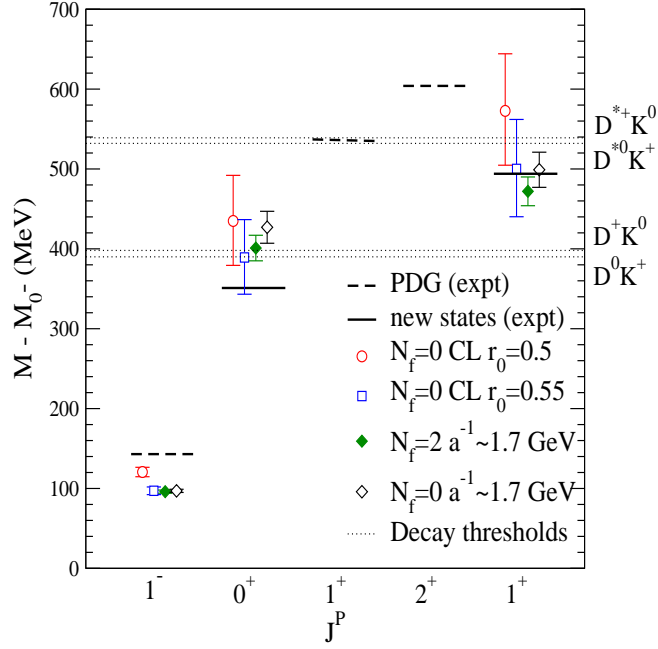} 
\caption{\label{fig:spectrum}Comparison of experimental results with the 
lattice determinations from the quenched continuum limit (open
symbols) and the $N_f=2$ data at fixed lattice spacing (solid symbols)
with $r_0=0.55$ fm. For comparison we also show the quenched result at
the same lattice spacing (open diamonds). Zero on the vertical scale
is set by the $D_s^+(1969)$ mass. Also plotted are the experimental
$DK$ and $D^{\star}K$ thresholds.}
\end{center}
\end{figure}

Shown in figure~\ref{fig:spectrum} and Table~\ref{tab:spectrum} is the
comparison of the lattice results along with the experimentally
measured spectrum. The dynamical results appear to have smaller error
bars and to be systematically higher than the quenched result. These
effects are due to extrapolating the quenched results and are absent
at fixed lattice spacing. The computed $1^--0^-$ splitting is too
small, a well known failing of the quenched
approximation~\cite{Mackenzie:1998kd,McNeile:2002uy}.  Evidently the
dynamical sea quark mass is too large to change this, as has been
observed before in the light hadron spectrum on the same
ensemble~\cite{DFLHS}. The lattice results for the $c\bar{s}$ $0^+$ and the
lightest $1^+$ mesons are consistent, albeit within large statistical
and systematic uncertainties, with the masses of the states discovered
recently by BABAR and CLEO. Although $u$ and $d$ sea-quark effects are not
yet properly included, these lattice results provide the most reliable
computation of the $c\bar{s}$ spectrum to date. Our errors are too
large to exclude exotic states based on potential models. However,
there is no evidence from our lattice QCD calculations that exotics
are required to explain the BABAR and CLEO discoveries.

\begin{acknowledgments}
The lattice data was generated on the Cray T3D and T3E systems at EPCC
supported by, EPSRC grant GR/K41663, PPARC grants GR/L22744 and 
PPA/G/S/1998/00777. We are grateful to the ULgrid project of the 
University of Liverpool for computer time.
The authors acknowledge support from EU 
grant HPRN-CT-2000-00145 Hadrons/LatticeQCD, and PPARC grants 
PPA/G/O/2000/00456, PPA/P/S/1998/00255(CMM),  PPA/N/S/2000/00217(RDK).

\end{acknowledgments}


\begin{table}
\caption{\label{tab:ensembles}Ensemble of gauge configurations. $\kappa_{\rm sea}=0$ denotes a quenched ensemble.}
\begin{ruledtabular}
\begin{tabular}{cccc}
$(\beta,\kappa_{\rm sea})$ & Volume & $a^{-1}$ GeV $r_0=0.55 $fm& number of configurations\\
\hline
$(6.2,0)$&$24^3\times 48$ &$2.64$ & $216$ \\
$(6.0,0)$&$16^3\times 48$ &$1.92$ & $302$ \\
$(5.93,0)$&$16^3\times 32$ &$1.70$ & $278$ \\
$(5.2,0.1350)$&$16^3\times 32$ &$1.70$ & $395$ \\
\end{tabular}
\end{ruledtabular}
\end{table}

\begin{table}
\caption{\label{tab:spectrum}Comparison of lattice results with experiment. Mass splittings from the $D_s^+(1969)$ Mass in MeV. CL denotes continuum limit,
$r_0=0.55$ fm is used unless otherwise noted.}
\begin{ruledtabular}
\begin{tabular}{ccccccc}
$J^P$& & experiment & $N_f=0$ CL & $N_f=0$ CL & $N_f=2$ &  $N_f=0$ \\ 
     & &     & $r_0=0.5$ fm & & $a^{-1}\sim 1.7$ GeV & $a^{-1}\sim 1.7$ GeV \\
\hline
$1^-$&        &$143\ \ \ \ $ & $121(6) $ & $  97(6) $ & $96(2)$   & $97(2)$\\
$0^+$&        &$351(1)$      & $435(57)$ & $389(47) $ & $401(16)$ & $427(20)$\\
$1^+$&$j=3/2$ &$576\ \ \ \ $ & $-$       & $-$        & $-$       & $-$ \\
$2^+$&        &$604\ \ \ \ $ & $-$       & $-$        & $-$       & $-$  \\
$1^+$&$j=1/2$ &$494(2)$      & $572(72)$ & $500(62) $ &$472(20)$  & $499(22)$\\
\end{tabular}
\end{ruledtabular}
\end{table}

\end{document}